\documentclass{article}
\usepackage{url}

\usepackage[utf8]{inputenc}
\usepackage[english]{babel}

\usepackage{amsmath}
\usepackage{amsfonts}
\usepackage{amssymb}

\usepackage{graphicx}

\usepackage{url}

\usepackage{color}
\usepackage{comment} 
\newif\ifshow
\showtrue
\showfalse
\ifshow

\else
\excludecomment{cmt}
\fi

\begin{document}

\title{A Process for the Evaluation of Node Embedding Methods in the Context of Node Classification}

\author{Christoph Martin\footnote{\small{\tt{cmartin@uni.leuphana.de}}} \and Meike Riebeling}

\date{%
    Institute of Information Systems\\ Leuphana University of L\"uneburg, 21335 L\"uneburg, Germany\\
    \vspace*{.5cm}
    \today
}

\maketitle              

\begin{abstract}
Node embedding methods find latent lower-dimensional representations which are used as features in machine learning models. In the last few years, these methods have become extremely popular as a replacement for manual feature engineering.

Since authors use various approaches for the evaluation of node embedding methods, existing studies can rarely be efficiently and accurately compared.
We address this issue by developing a process for a fair and objective evaluation of node embedding procedures w.r.t. node classification. This process supports researchers and practitioners to compare new and existing methods in a reproducible way.

We apply this process to four popular node embedding methods and make valuable observations. 
With an appropriate combination of hyperparameters, good performance
can be achieved even with embeddings of lower dimensions, which is positive for
the run times of the downstream machine learning task and the embedding algorithm. Multiple hyperparameter combinations yield similar performance.
Thus, no extensive, time-consuming search is required to achieve reasonable performance in most cases.

\end{abstract}

\section{Introduction}

Networks are used to model phenomenons in various domains such as social relations, molecular graphs, biological structures, or recommender systems. Networks represent the relations (edges) between different entities (nodes). Social networks contain information about individuals or communities and the dynamics among them. This information can, for example, be used for segmentation or recommendation tasks.
Networks capture not only social relationships, but also citations, biological information, or knowledge relations~\cite{Newman2003}. Developing and experimenting with methods that leverage the information captured by these networks are important endeavors in business and research communities~\cite{yang2015TADW,2018Survey}.

In various fields, network data are to be used as input for machine learning models.
This poses the challenge that network data must first be transformed in order to serve as features.
Traditionally, handcrafted features have been created to represent the nodes.
This type of feature engineering, however, has considerable weaknesses. It is very time-consuming on the one hand and, on the other hand, the handcrafted features can often not be reused~\cite{Hamilton2017}.
Node embeddings map the nodes of a graph to a lower-dimensional vector which can subsequently be used as input for other machine learning techniques.
However, due to the particular data structure of a network, the quality of network embeddings depends on preserving the structural properties of a graph while incorporating node attributes.
This can be difficult as the structural similarity of nodes can either be portrayed as nodes close to each other or as nodes with similar roles in the network, node embeddings have to respect local and global node similarities together~\cite{Cai2018,2018Survey,Goyal2018}.

Node embedding methods have enormous potential, thus this area continues to be a highly active field of research.
In recent years, several surveys have been published, which summarize the progress made in this area and address the comparison and categorization of node embedding methods~\cite{Hamilton2017,Goyal2018,2018Survey,Cai2018}.
Due to the popularity of embedding methods, a unified way to compare them has become increasingly important. Methods proposed by existing studies can rarely be compared to each other since authors use different approaches to evaluate node embeddings.

We address this issue by developing a process (Section~\ref{sec:devel_framework}) for a fair and objective evaluation of node embedding procedures w.r.t. node classification.
Building on and extending existing work~\cite{Goyal2018,2018Survey,Khosla2019,goyal2019benchmarks}, we explicitly address the choice of the hyperparameters in the process presented here, under consideration of the downstream machine learning task, in this case node classification.
This process supports researchers to compare new and existing methods in a reproducible way. Furthermore, end users can use this process to find the optimal method for the particular use case.

In the case study in Section~\ref{sec:casestudy}, we apply the process to four popular node embedding methods and make valuable observations, especially for practitioners.
The default hyperparameters for node embedding procedures are generally not a good choice.
With an appropriate combination of hyperparameters, good performance can be achieved even with embeddings of lower dimensions, which is positive for the run times of the downstream machine learning task.
Multiple hyperparameter combinations yield similar performance; hence usually there is no extensive, time-consuming search required to achieve reasonable performance.

\section{Node Embeddings}

Let $G$ be a graph on $N$ nodes with vertex set $V(G) = \{v_1, v_2, \ldots, v_N\}$. Node embeddings are d-dimensional representations of the nodes in $G$; usually, these are lower-dimensional (i.e., $d \ll N$). These embeddings are commonly used as input for machine learning algorithms. Node embedding methods have the objective to find such a mapping $f: V(G) \rightarrow \mathbb{R}^{d}$, where nodes which are ``similar'' to each other in the graph also ``similar'' to each other in the vector space.
The definition of similarity differs between methods.
In the literature, the terms graph embedding or network relational learning are also used for this purpose~\cite{Perozzi2014,2018Survey,Hamilton2017}.

In our case study (Section~\ref{sec:casestudy}), we use the following four frequently cited and widely used node embedding methods: node2vec \cite{Grover2016}, GraRep \cite{Cao2015}, Deep Network Graph Representation (DNGR) \cite{cao2016DNGR}, and Large-scale Information Network Embedding (LINE) \cite{tang2015line}. We use the implementations provided by~\cite{GIT_gem,OPNENE_GIT}.

\section{A Process for the Comparison of Node Embedding Methods}
\label{sec:devel_framework}

In this section, we develop the evaluation process for node embedding methods.
This process enables researchers and practitioners to perform a fair and objective evaluation of node embedding procedures. We present this process for two main reasons. The first is to compare new and existing methods in a reproducible way.  Furthermore, it helps end users to find the optimal method for the particular use case.
We start by arguing why the procedure for selecting hyperparameters cannot easily be transferred from previous machine learning methods to node embedding learning. Then we propose an approach and integrate it with the process.

The evaluation of algorithms and methods is an essential part of machine learning and network analysis research \cite{caruana2006empirical,daelemans2002evaluationNLP}. Particularly, algorithm selection is a widely discussed topic and an essential part of the application of machine learning algorithms in practice. This is due to the fact that there is not one single method optimal for all problem settings~\cite{kou2012evaluation,nofreelunch}.

Essential components of evaluation experiments in machine learning are the data set, feature selection, feature representation, and hyperparameter settings~\cite{daelemans2002evaluationNLP}. The components of an evaluation process for node embeddings are slightly different. The data set and the hyperparameter settings can be transferred to node embeddings as essential components of the evaluation \cite{2018Survey}. However, the feature selection process and the data representation have to be altered. Node embedding methods naturally take a network and the contained information as feature input, essentially making the step of feature selection unnecessary. 
The necessary representation of the network might differ between algorithms, hence the data representation is implied by choice of the embedding method.

Node embedding methods constitute an unsupervised problem setting traditionally; semi-supervised methods also exist (e.g., \cite{Kipf2017,Shchur2018}), but these are not addressed in this paper.
An application task is, therefore, necessary to evaluate the quality of node embeddings and is thus an essential component of the process.
In summary, the core components of the process are the network data, the application task, the evaluation metric, and the hyperparameter configuration.

\paragraph*{Network data}
The choice of the network data depends on the setting in which the process is applied and the node embedding methods considered.
Practitioners who are looking for the best method for their particular application should use data that is close to the production data.
For the comparative evaluation of new and existing embedding methods, in the interests of reproducibility, we recommend using publicly available networks of different size and structure. These may be, for example, the data sets used in the case study in Section~\ref{sec:casestudy}.

\paragraph*{Application task and evaluation metric}
The most  popular application task is node classification, which is often applied when presenting a new embedding method. Classification aims at finding class labels for each node. The vector representation serves as feature input for a classifier~\cite{Cai2018,Goyal2018}. Training a classifier requires training data, which means that labels have to be available at least for a part of the network. Common evaluation metrics in this context include F$_{\text{1}}$-score, precision, recall, or accuracy.
We propose to use the F$_{\text{1}}$-score since it takes precision and recall into account,
$\text{F}_{\text{1}}\text{-score} = \frac{2 \cdot \text{precision} \cdot \text{recall} }{\text{precision} + \text{recall}}$. 
In the case of multi-class and multi-label classification problems, we use the macro and micro variant of the F$_{\text{1}}$-score. Here the classes, respectively, the individual observations, are weighted equally~\cite{powers2011evaluation}.

For the classification task, we propose to use two popular and often used algorithms in machine learning: a logistic regression model (one-vs-rest classification for the multi-label model) and a random forest model. The regression model because of the frequent usage in the evaluation of embedding methods. The random forest is a widely used model in practical machine learning applications. Nevertheless, it is usually not applied in node embedding research. 
Therefore we suggest to use it in this context because it is very flexible and leads to good results on different data sets~\cite{JMLR:v15:delgado14a}.

\paragraph*{Hyperparameter configuration}

\begin{figure}[t]
\centering
\includegraphics[width=\textwidth]{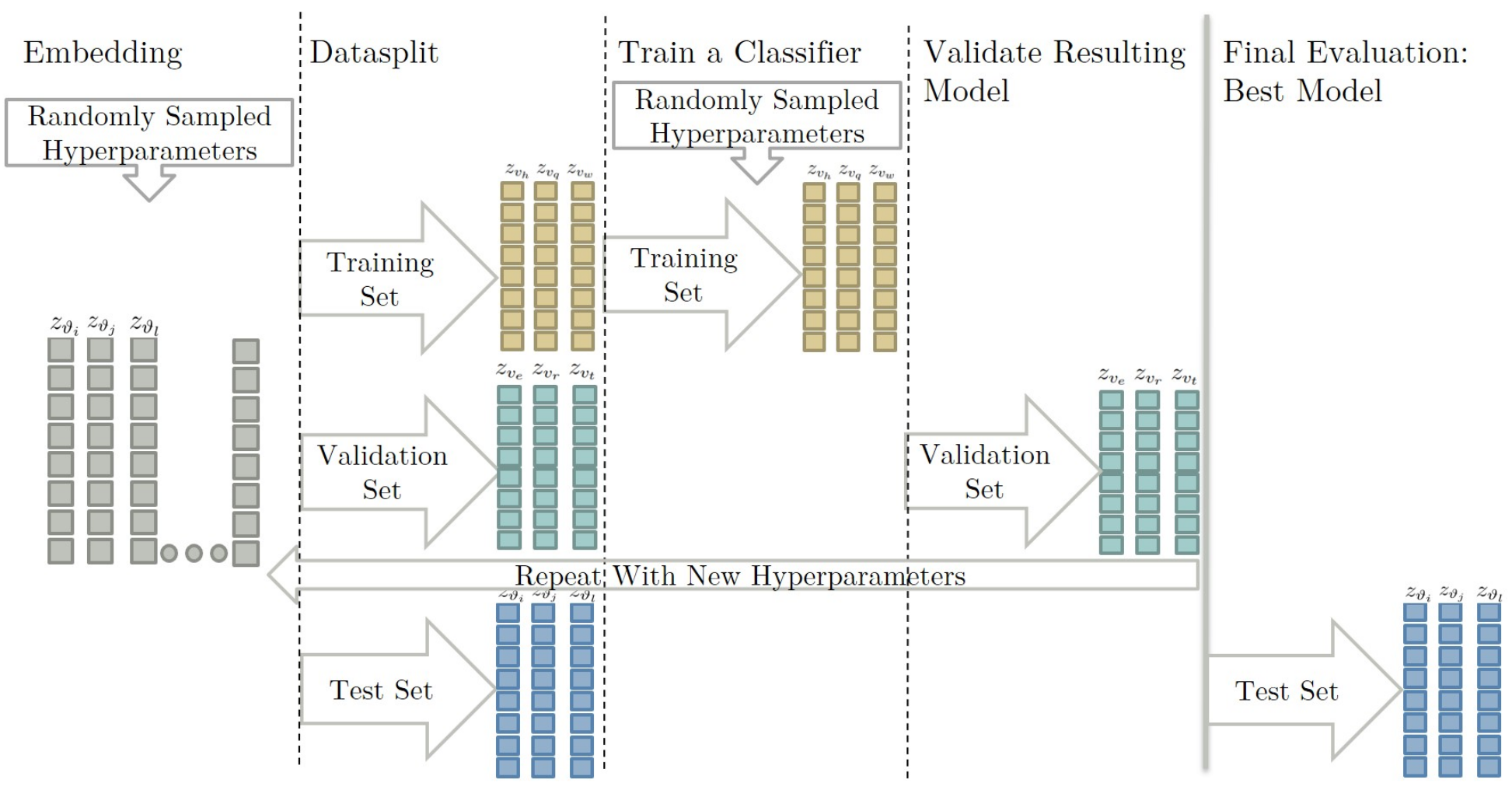}
\caption{\textbf{Setup for the hyperparameter selection} The
embedding algorithm is applied to the whole network resulting in a vector representation
marked by gray squares on the left side which is subsequently divided
into three splits: training set (yellow), validation set (turquoise) and test set (blue). The training set is used to train the classifier, the validation set is used in that last part of the hyperparameter selection to validate
the performance of the combined model.
The test set is not used during the hyperparameter
selection. After repeating this process several times, the best hyperparameter combination is selected for the final model, which is evaluated on the test set (final step on the right separated by the bold line).}
\label{fig:HyperparameterSelection}
\end{figure}

The selection of the best hyperparameters is a debated topic in research.
The impact of different tuning parameters on each other and how they affect the performance is only poorly understood \cite{hyperband}.
In practice, a widely used method to find a set of hyperparameters is random search, where the search space of hyperparameters is randomly explored and evaluated. Begstra and Bengio \cite{bergstra2012randomSearch} showed that this type of search leads to equally good or even superior models, compared to grid search, while only a fraction of the time is needed.

In addition to the way the hyperparameter selection is performed, the data utilized for tuning is an important topic. Usually, in machine learning data is split into a test, training, and validation set, in which the test set is only used once for the final validation. The training of the algorithm is performed on the training set with a subsequent evaluation of the performance using the validation set \cite{james2013introductionSTATISTICALLEARNING}. For network embedding procedures, this is not possible. Splitting the network data into different sub-graphs would significantly alter the results of the embedding methods as they rely on representing the whole graph mirroring the structural context information of a node and its position in the whole network. Only using part of the network for an embedding would lead to a completely different representation with important context information missing. 
The proposed solution for the described challenges is a combined tuning of hyperparameters of the embedding and the subsequent application algorithm. The application task serves as the basis for the performance evaluation governing the hyperparameter selection. As shown in Figure \ref{fig:HyperparameterSelection}, the  representation for the whole graph is learned, whereas only part of this data is used in the application task (for example, classification) to evaluate the hyperparameter selection. For both algorithms -- the embedding algorithm and the classification algorithm -- hyperparameters are selected randomly. This process is repeated several times. Finally, the best model combination using the best hyperparameters for both algorithms is picked and evaluated on the test set.

\section{Case Study for the Comparison Process}
\label{sec:casestudy}

In this section, we utilize the process developed in Section~\ref{sec:devel_framework} to compare four frequently cited and widely used node embedding methods: node2vec, GraRep, LINE, and DNGR. Especially, we are interested in the impact of the number of dimensions and the amount of training data used on the performance in the domain of node classification.

We use data sets with varying characteristics (i.e., directed and undirected as well as binary, multi-class, and multi-label classification)  to get an understanding of how embedding procedures behave under different conditions. Table~\ref{tab:overviewDatasets}  lists basic statistics about these networks.
For training and model selection, we use $50\%$ for the training set and $25\%$ for the validation set and test set.
For the second part of the experiment, where we analyze the impact of varying amounts of training data, we use $10\%, 20\%, \ldots, 100\%$ of the training data. All of  these values refer to the node embedding vectors.

\begin{table}[hbtp]
\caption{Summary of networks used in case study.}
\label{tab:overviewDatasets}
\centering

\setlength{\tabcolsep}{.5em} 
\begin{tabular}{lrrrrr}
\hline
Network      & Nodes  & Edges   & Directionality & \# Labels         & Source                                \\ \hline
Moreno Blogs & 1,224  & 19,025  & directed       & 2 (binary)       & \cite{Moreon_Blog_2005political}      \\
CiteSeer     & 3,312  & 4,660   & directed       & 6 (multi-class)  & \cite{CiteSeer2005link}               \\
Facebook     & 4,039  & 88,234  & undirected     & 4 (multi-class)  & \cite{facebook_article_socialCircles} \\
BlogCatalog  & 10,312 & 333,983 & undirected     & 39 (multi-label) & \cite{tang2009relational}             \\ \hline
\end{tabular}
\end{table}

\paragraph{Overall results}

The performance of the embedding methods w.r.t. the different classifiers and measures are listed in Table~\ref{tab:resultsExperimentsBothScores}.
The scores for the logistic regression scenarios reveal that most of the tested algorithms perform similar across the networks. The highest score for the BlogCatalog network is 0.35, which was reached by node2vec. LINE and GraRep reach equal scores of 0.34 on that network. For Facebook, the scores are even closer together, the values vary between 0.45 and 0.52. The same trend can be found in the results of the Moreno network. For the Moreno network, the score of LINE, GraRep, DNGR, and node2vec are the same with 0.95. The best scores for CiteSeer range from 0.53 to 0.57. Only the deep learning-based method yield worse results, DNGR does not work well with a score of 0.25.  Overall, the results indicate that very similar scores can be reached across different methods.
The observed performance of node2vec, LINE, and GraRep on the BlogCatalog data set are in line with the results reported in the literature. For GraRep and node2vec, evaluation  experiments were also conducted using a one-vs-rest logistic regression \cite{Cao2015,Grover2016}. Moreover, in \cite{Cao2015}, LINE was included as a baseline. For all three networks, the performance was around 0.4; the slightly lower performance observed in this paper might be explained by the use of only 50\% of the networks for training, due to the data split in training, validation and test set explained above.

\begin{table}[htbp]
\caption{Results for the experiments in the case study.}
\label{tab:resultsExperimentsBothScores}
\centering

\setlength{\tabcolsep}{.2em}

\begin{tabular}{lllrrrr}
\hline
                     &                     & Network   & BlogC. & CiteSeer & Facebook & Moreno \\
Score                & Classifier          & Embedding &             &          &          &        \\ \hline
Macro F$_{\text{1}}$ & Random forest       & DNGR      & 0.020       & 0.180    & 0.434    & 0.941  \\
                     &                     & GraRep    & 0.111       & 0.555    & 0.489    & 0.944  \\
                     &                     & LINE      & 0.020       & 0.240    & 0.458    & 0.954  \\
                     &                     & node2vec  & 0.032       & 0.505    & 0.456    & 0.951  \\
                     & Log. regression & DNGR      & 0.068       & 0.153    & 0.485    & 0.954  \\
                     &                     & GraRep    & 0.181       & 0.514    & 0.505    & 0.951  \\
                     &                     & LINE      & 0.195       & 0.469    & 0.427    & 0.948  \\
                     &                     & node2vec  & 0.212       & 0.493    & 0.412    & 0.954  \\
Micro F$_{\text{1}}$ & Random forest       & DNGR      & 0.052       & 0.266    & 0.450    & 0.941  \\
                     &                     & GraRep    & 0.244       & 0.607    & 0.505    & 0.944  \\
                     &                     & LINE      & 0.054       & 0.273    & 0.507    & 0.954  \\
                     &                     & node2vec  & 0.088       & 0.563    & 0.514    & 0.951  \\
                     & Log. regression & DNGR      & 0.182       & 0.248    & 0.507    & 0.954  \\
                     &                     & GraRep    & 0.342       & 0.574    & 0.525    & 0.951  \\
                     &                     & LINE      & 0.339       & 0.536    & 0.450    & 0.948  \\
                     &                     & node2vec  & 0.354       & 0.534    & 0.496    & 0.954  \\
\multicolumn{3}{l}{Most frequent label}                & 0.090       & 0.212    & 0.336    & 0.520  \\ \hline
\end{tabular}

\end{table}

\paragraph{Analysis of the number of dimensions}

\begin{figure}
 \caption{Impact of the dimensionality on the performance for the Facebook network.}
 \label{fig:resultsDimensions}
 \includegraphics[width=1\textwidth]{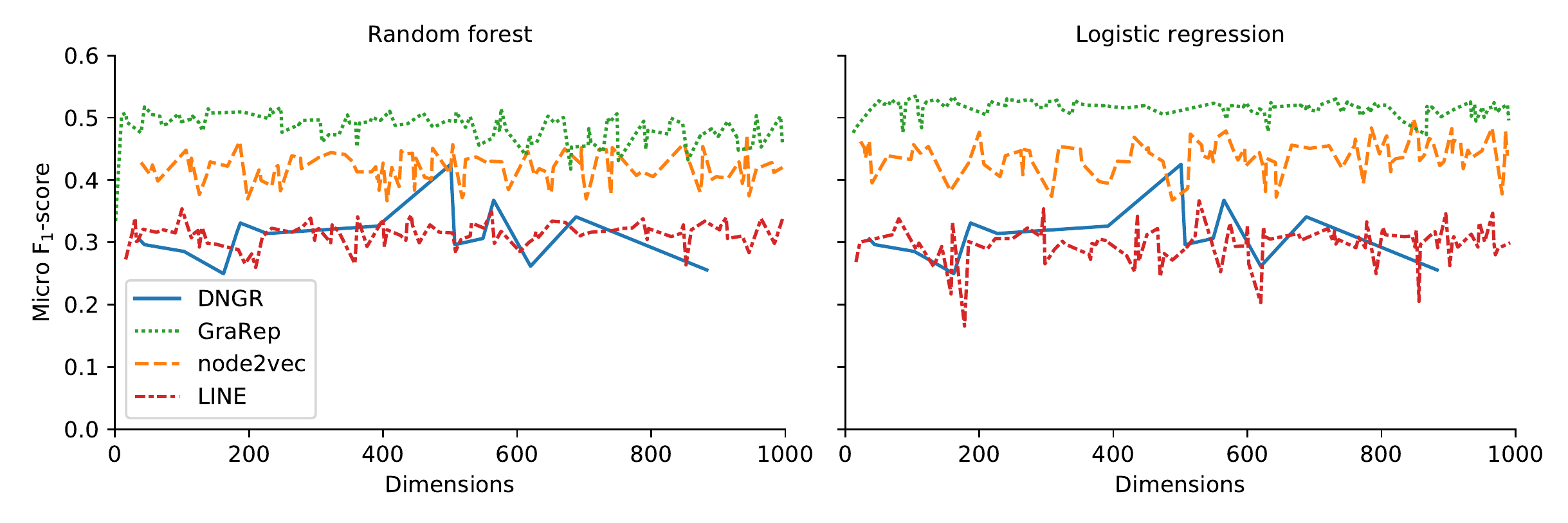}
 \end{figure}

The dimensionality of the embedding is the only hyperparameter shared by all node embedding methods. The performance of embedding algorithms should, intuitively,  increase with an increasing number of dimension until reaching a plateau where no substantial improvement of performance happens with increasing dimensionality.
Grover and Leskovec~\cite{Grover2016} observed this behavior for the node2vec algorithm. Experimenting with the number of dimensions resulted in a saturation of performance improvement at a dimension of around 100.
Similar results are reported by Wang et al. \cite{wang2016SDNE}. They noticed a decline in performance after saturation at about 100.
For some algorithms like GraRep, little influence of the dimensionality on performance was observed. The reported relation between dimension and performance is almost steady, with a slight decrease after 64 dimensions~\cite{Cao2015}.
 In Figure~\ref{fig:resultsDimensions}, the performance depending on the dimension for the case of the Facebook network is shown.
 These results indicate that higher dimensions do not necessarily lead to better performance.
 This behavior also occurs for the other networks.
 However, analyzing the performance with different dimensions lead to high variances.
 The reason might lie in the high amount of different hyperparameter combinations since the performance is not only dependent on the dimension but on the combination of parameters picked.
 Nonetheless, the findings suggest that in combination with the right hyperparameters, small dimensions are sufficient to reach scores, that are comparable to the performance with higher dimensions.
 The results highlight the influence of all hyperparameters on each other.
 Therefore, the optimal performance of an embedding method depends on all hyperparameters, the network, and the application task.
 Moreover, the results suggest that equally well results can be reached with many different hyperparameter combinations, indicating that a reasonable performance can be reached without an extensive hyperparameter search. 
 This may also explain the difference between our results to the above. We consider the performance of the final application task (node classification) when finding embeddings.

\paragraph{Hyperparameter for node2vec}

A more detailed analysis of the hyperparameter for node2vec is listed in Table~\ref{tab:node2vecHyperparameter}.
The results lead to the conclusion that the best hyperparameter combination depends on the network and the application task. In the case of the BlogCatalog data set, there are also apparent differences between the two classification algorithms: The hyperparameter search leads the algorithm towards different learning strategies. The values for the sampling parameters $p$ and $q$ are 2 and 0.25 in the random forest case and 2 and 1 for the logistic regression. Thus for the Blog Catalog network, the random forest benefits from a depth-first sampling strategy preferring nodes further away from the source node, whereas the sampling strategy for the logistic regression is not biased towards one sampling strategy. The parameter $p$ is 2 for both classification cases. Hence, the likelihood of revisiting a node is low.

\begin{table}[htbp]
\caption{Results for the hyperparameter analysis of node2vec.}
\label{tab:node2vecHyperparameter}

\setlength{\tabcolsep}{.1em} 
\centering
\begin{tabular}{lrrrr|rrrr}
\hline
& \multicolumn{4}{c}{Random forest}     & \multicolumn{4}{c}{Logistic regression} \\
& Blog  & Facebook & Cite. & Moreno & Blog   & Facebook  & Cite. & Moreno \\
\hline                      
Micro F$_{\text{1}}$-score                 & 0.04      & 0.47     & 0.27  & 0.95   & 0.22       & 0.5       & 0.35  & 0.95   \\
Dimension             & 74        & 943      & 103   & 733    & 197        & 848       & 245   & 600    \\
Return parameter: $p$ & 2         & 0.5      & 0.25  & 1      & 2          & 2         & 0.75  & 2      \\
In-out parameter: $q$ & 0.25      & 0.5      & 0.5   & 4      & 1          & 0.5       & 0.25  & 0.5    \\
Number of walks: $l$   & 25        & 42       & 40    & 20     & 33         & 5         & 48    & 39    \\
Walk length: $k$      & 70        & 56       & 11    & 45     & 46         & 28        & 11    & 24     \\
\hline
\end{tabular}

\end{table}

In the paper introducing node2vec,  experiments were also conducted on the BlogCatalog network. The authors described an increase in performance with small values for p and q. The results of the presented experiments suggest higher values for p and smaller values for q. Differences in the findings for the return parameter p are probably due to the variants in the remaining hyperparameters. Grover and Leskovec \cite{Grover2016} used default values for all remaining hyperparameters and only experimented with the values for p and q. The findings of the random search suggest a strong effect of the interaction between the parameters. Even though a lower $p$ is optimal in the case of default parameters, a higher $p$ --~leading to node sequences containing samples further away from the source~-- leads to better results, when combined with more and longer walks. 
These observations highlight the importance of tuning the hyperparameters of node embeddings based on the application task instead of simply using the default parameters.

\paragraph{Impact of \% training data on the performance}

\begin{figure}[htbp]
\centering
\includegraphics[width=1\textwidth]{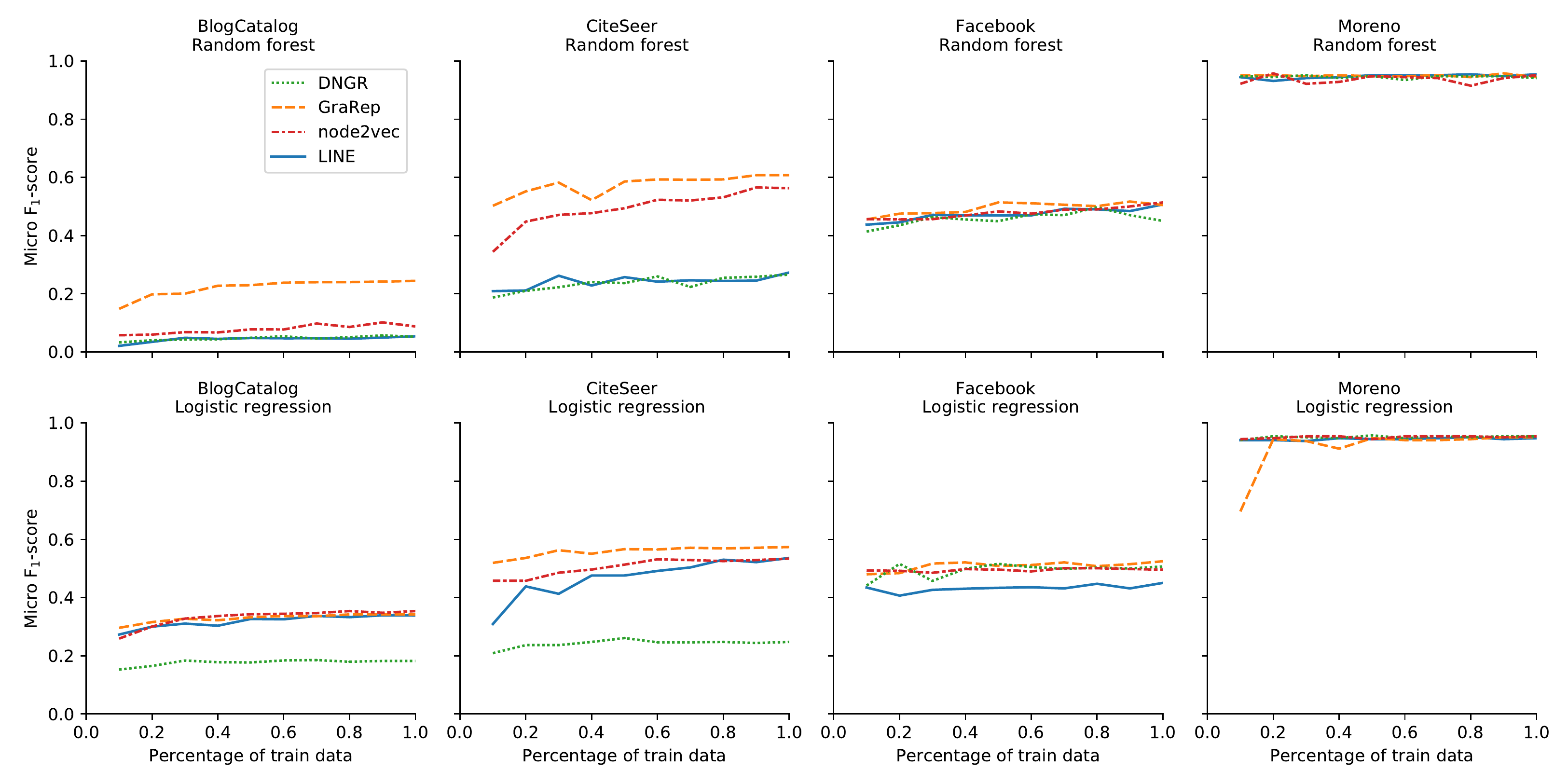}
\caption{Results for the impact of an increasing amount of training data on the classification performance.}
\label{fig:increasingTraindataRFundLog}
\end{figure}

Figure~\ref{fig:increasingTraindataRFundLog} shows the impact of increasing the amount of training data on the performance, the overall impact is small. The behavior of the curves, however, shows that with small ratios, an increase in the amount of training data has a high impact on the performance. At some point, an increase in the data only leads to a small improvement. 
As an example, for both classification methods, the performance for the Moreno network reaches a peak in performance increase at around 20\% of the training data. After that point, the impact on performance is relatively low. Similarly, the score for the embedding methods on the BlogCatalog network are increasing until a ratio of 0.2 to 0.3, the performance score of node2vec in the logistic regression scenario is 0.26 with 10\% of the data. However, with 30\% of the data, the score is already 0.33. There is only little improvement thereafter as the best score is 0.35. This is consistent with previous studies that showed that the performance of node2vec shows large improvements until 30\% after that, the increase in performance is small \cite{Goyal2018}.
For CiteSeer differences between the random forest and the logistic regression scenario can be observed. In the case of the random forest combined with GraRep and node2vec there is a substantial increase in performance. The starting value of node2vec, for example, is 0.34, whereas the best performance is 0.56. However, in the logistic regression scenario, the difference is only 0.05, which is consistent with a similar experiment conducted by \cite{2018Survey}, who compared the results using 5\% and 50\% of the whole network data and found an increase of 0.08 points for CiteSeer. The reason for these differences in the two application scenarios is not apparent. However, it might be because the random forest needs more labeled observations to separate them efficiently. The CiteSeer network has many labels with only a few observations. Therefore, a small amount of data might lead to an underrepresentation of training data for some labels.

\section{Conclusions}

Recently, node embeddings became popular as an alternative to handcrafted feature engineering~\cite{Hamilton2017}.
In this paper, we proposed a process for the comparison of node embedding methods w.r.t. node classification.
This process enables researchers and practitioners to perform a fair and objective evaluation of node embedding procedures and helps end users to find the optimal method for the particular use case.

Moreover, in a case study, we applied this process to four popular node embedding methods.
These experiments showed that the introduced process provides a foundation for a standardized evaluation of node embedding methods. Additionally, we made 
valuable observations, especially for practitioners:
The default parameters for node embedding procedures are generally not a good choice. We analyzed this in detail for node2vec.
Analyzing the impact of the dimensionality of the embeddings, we noticed that
the appropriate combination of hyperparameters yields good performance with a lower number of dimensions, which is positive for the run times of the downstream machine learning task and the embedding algorithm. We also observed that multiple hyperparameter combinations yield similar performance. Hence there no extensive, time-consuming search required to achieve reasonable performance.

Although the proposed process provides a robust foundation for the comparison of node embedding methods, there are some aspects which should be addressed by future research. For example, the application task link prediction. It would be particularly interesting to understand how the procedure has to be adjusted differently for missing and future link prediction.
A comprehensive comparison of semi-supervised methods would also be of interest.

\bibliographystyle{acm}
\bibliography{myLibLink.bib}

\end{document}